\documentclass {llncs}
\usepackage {qtree}
\usepackage {hyperref}
\usepackage {amssymb}
\usepackage[OT1,T1]{fontenc}
\usepackage{verbatim}
\newcommand{\gl}{\guilsinglleft}
\newcommand{\gr}{\guilsinglright}
\newcommand{\term}[1]{\texttt{\small{#1}}}
\newcommand{\nterm}[1]{\textrm{\gl #1\gr}}

\newcommand{\tv}{{\Large $\tau_\nu$} }

\newcommand{\tline}{\begin{tabular*}{\textwidth}{l}\hline}
\newcommand{\bline}{\hline\end{tabular*}}

\newtheorem{defn}{Definition}

\newcommand{\olabel}[1]{\textbf{{#1}}}
\begin {document}

\title {Enhancing the Alloy Analyzer with Patterns of Analysis}
\author {William Heaven and Alessandra Russo}
\institute {Department of Computing, Imperial College London, SW7 2AZ\\\textsf{\{william.heaven, ar3\}@doc.ic.ac.uk}}
\maketitle
\vspace*{-5pt}
\begin{abstract}Formal techniques have been shown to be useful in the development of correct software. But the level of expertise required of practitioners of these techniques prohibits their widespread adoption. Formal techniques need to be tailored to the commercial software developer. Alloy is a lightweight specification language supported by the Alloy Analyzer (AA), a tool based on off-the-shelf SAT technology. The tool allows a user to check interactively whether given properties are consistent or valid with respect to a high-level specification, providing an environment in which the correctness of such a specification may be established. However, Alloy is not particularly suited to expressing program specifications and the feedback provided by AA can be misleading where the specification under analysis or the property being checked contains inconsistencies. In this paper, we address these two shortcomings. Firstly, we present a lightweight language called \emph{Loy}, tailored to the specification of object-oriented programs. An encoding of Loy into Alloy is provided so that AA can be used for automated analysis of Loy program specifications. Secondly, we present some \emph{patterns of analysis} that guide a developer through the analysis of a Loy specification in order to establish its correctness before implementation. \end{abstract}

\section {Introduction}

Since the late 60s and the presentation of a logic for programs \cite{floyd67,hoare69}, a large field of Computer Science research has been concerned with the application of formal techniques to software development in order to provide some  guarantee that the software will behave as expected \cite{apt81}. Commercial pressure to produce higher-quality software is always increasing \cite{clarke96,observer05}. But despite the considerable advances made in this area from a research point of view, very few of these techniques are applied today in industry. On the whole, a combination of the level of expertise required of practitioners of these techniques and the apparent extra costs they impose on software development have made their adoption commercially prohibitive. Software developers favour so-called \emph{lightweight} techniques that provide more immediate returns and that sit more comfortably with the activity of implementation \cite{jackson96}. 
\vspace*{0pt}\\
\hspace*{\parindent} 
Several lightweight specification languages supported by an array of tools have been proposed. For example, JML \cite{leavens98} and Spec$\sharp$ \cite{barnett04} allow implementers to annotate source files with formal specifications. These specifications can be used by tools to check the correctness of an implementation, statically or at runtime \cite{burdy05,barnett04}. Such tools are particularly suited to catching null-pointer and index-out-of-bounds exceptions or violations of invariants and method specifications in an implementation. However, they do not support a direct check of the consistency of a specification and inconsistency sometimes becomes manifest only through the unexpected results of analysis. For example, ESC/Java2, a tool that statically checks a Java implementation against a JML specification, will always pass a method body when checking it against its specification if the precondition of the method is unsatisfiable, since the tool does not check the bodies of methods that do not satisfy their precondition. To avoid misleading results and, of course, since an inconsistent specification has no possible implementation, a developer should be able to establish the consistency of a specification before implementation is attempted.
\vspace*{0pt}\\
\hspace*{\parindent} 
Alloy \cite{jackson04}, another lightweight specification language, is supported by the Alloy Analyzer (AA) \cite{jackson00}, a tool based on off-the-shelf SAT technology. The tool allows a user to check whether given properties are consistent or valid with respect to a specification, providing an environment in which the correctness of a specification may be established. However, the environment has two shortcomings for the software developer: Alloy is not specifically suited to expressing \emph{program} specifications and feedback from AA can be misleading when the specification under analysis or property being checked contains inconsistencies.
\vspace*{0pt}\\
\hspace*{\parindent} 
Firstly, Alloy does not have an implicit concept of state. For example, a relation between the before and after value of a field \emph{f} in a JML method specification can be specified in terms of a relation between the variables \emph{$\backslash$old(f)} and \emph{f}. But Alloy has no comparable implicit means to denote the before and after value of a field. In Alloy, this relation would have to be expressed by explicitly constructing the necessary constraints between two intrinsically unrelated variables \emph{f} and \emph{f~$'$}, say. In practice, specifying these extra constraints is a complicated task and can lead to an overly cluttered specification. 
\vspace*{0pt}\\
\hspace*{\parindent} 
Secondly, feedback from AA can be misleading when the specification under analysis or the property being checked contain inconsistencies. If a user of the tool is unaware of an existing inconsistency, accepting such feedback at face value can lead to further error. 
For example, consider the small Alloy specification shown below. 
\vspace{8pt}
\\
\hspace*{20pt}\term{sig Project \{ \} }
\\
\hspace*{20pt}\term{sig Employee \{ project : Project \} }
\\
\hspace*{20pt}\term{sig Pool extends Employee \{ \} \{ no project \} }
\\
\hspace*{20pt}\term{fact \{ some Pool \} }
\vspace{8pt}
\\
The specification represents a scenario in which employees are assigned pro\-jects and some employees belong to a pool. Employees belonging to the pool have no assigned project. However, there is an inconsistency. The Alloy declaration \emph{project : Project} indicates that each employee has a project but employees belonging to the pool are specified to have no project. Since there must be at least one employee in the pool (because of the constraint imposed by the \emph{fact} paragraph) the specification is unsatisfiable. Now, if AA were asked whether or not some property were consistent with the specification, such as whether there could be some employees not in the pool ---
\vspace{8pt}
\\
\hspace*{20pt}\term{pred PropertyTest () \{ some e}~\term{:}~\term{Employee | e not in Pool \} }
\vspace{8pt}
\\
--- the tool would be unable to instantiate a model of the specification that satisfies this property because it is unable to instantiate a model of the specification itself. But AA simply suggests that the property may be inconsistent. If a user has no reason to suspect the fault of the specification, not realising that \emph{nothing} is consistent with it, the tested property will be rejected. In short, simply informing a user that a model cannot be found, without warning of possible mis-specification, is not adequate feedback.
 \vspace*{0pt}\\
\hspace*{\parindent} 
This paper addresses these two shortcomings. We present a lightweight language called \emph{Loy}, tailored to the specification of object-oriented programs. In particular, Loy allows the specification of before and after state values of fields in method specifications. We implement analysis support for this language by encoding Loy specifications into Alloy, invoking AA on the encodings and feeding back results in terms of the original Loy specification. But, most importantly, we show how the checking of specifications written in this language can be supported by \emph{patterns of analysis} that guide specifiers in questioning feedback from AA. Section~\ref{sec:alloy} gives an overview of how AA checks an Alloy specification. Section~\ref{sec:loy} introduces the language Loy. Section~\ref{sec:patterns} presents patterns of analysis for Loy specifications and Section~\ref{sec:encoding} discusses an encoding of Loy into Alloy that allows automation of the patterns to be implemented on top of AA. Section~\ref{sec:related} discusses related work and Section~\ref{sec:conclusion} concludes. 
\vspace*{-5pt}
\section {The Alloy Analyzer}\label{sec:alloy}
\vspace*{-5pt}
An Alloy specification consists of a set of \emph{signature paragraphs}, a set of \emph{fact paragraphs}, a set of \emph{predicate paragraphs}, a set of \emph{assert paragraphs} and a set of \emph{function paragraphs}. A signature paragraph declares a type and contains a set of field declarations for instances of that type. Semantically, a type is treated as a set, or \emph{domain}, and its instances as the elements of that domain. A field is treated as a relation between domains. The constraints introduced implicitly by the signature declarations together with the constraints enclosed by the fact paragraphs constitute a set of \emph{core constraints}, which are expected to be satisfied by all models of the specification. 
Predicate paragraphs enclose a set of formulae that can be tested for \emph{consistency} with respect to a specification and assert paragraphs enclose a set of formulae that can be tested for \emph{validity} with respect to a specification. 
Function paragraphs are essentially macro-expanded expressions and not relevant to the considerations of satisfiability in this paper. They are therefore not discussed further. 
\vspace*{0pt}\\
\hspace*{\parindent} 
A \emph{binding} is a function that maps variables of the specification according to their types to elements of the domains declared by the signature paragraphs. The semantics of a  specification can be given by a set 
$\{C, P_1, \dots, P_n, A_1, \dots, A_m\}$
where $C$ is a set of bindings associated with the core constraints, $P_i$ is a set of bindings associated with the constraints of a predicate paragraph $i$ and $A_j$ is a set of bindings associated with the constraints of an assert paragraph $j$. For a set of constraints $c$, $\vartheta \vDash c$ means that binding $\vartheta$ \emph{satisfies} the constraints of $c$.
\begin{defn}[Bindings for the core constraints]
Let $\gamma$ be the set of core constraints. $C = \{ \vartheta \mid \vartheta \vDash \gamma \}$ is the set of bindings
associated with the constraints of $\gamma$. 
The bindings of $C$ bind variables of the specification such that the constraints of $\gamma$ are satisfied. 
\end{defn}
\hspace*{\parindent}The constraints of a predicate paragraph are expected to be consistent with respect to the core constraints. The bindings of $P_i$ are those bindings that satisfy the core constraints and the constraints of predicate paragraph $i$ of the specification.
\begin{defn}[Bindings for a predicate paragraph]
Let $\gamma$ be the set of core constraints and let $i$ be a predicate paragraph. $P_i = \{ \vartheta \mid \vartheta \vDash \gamma \land i\}$ is a set of bindings that bind variables of the specification and the variables of $i$ (including the parameters of $i$) such that the constraints of $\gamma$ and $i$ are satisfied. 
\end{defn}
\hspace*{\parindent}The constraints of an assert paragraph are expected to be valid with respect to the core constraints. That is, there 
should be no binding that satisfies the core constraints but not the constraints of an assert paragraph. The bindings of $A_j$ are those bindings that satisfy the core constraints but not the constraints of assert paragraph $j$ of the specification. $A_j$ is expected to be the empty set. 
\begin{defn}[Bindings for an assert paragraph]
Let $\gamma$ be the set of core constraints and let $j$ be an assert paragraph. $A_j = \{ \vartheta \mid \vartheta \vDash \gamma \land \lnot j$\} is a set of bindings that bind variables of the specification and the variables of $j$ such that the constraints of $\gamma$ and $\lnot j$ are satisfied. 
\end{defn}
\hspace*{\parindent}In practice, AA searches for a model of a specification within a given \emph{scope}. That is, given a scope of 3, the tool searches for a model that satisfies a set of constraints such that there are at most three instances of each type in the model. Therefore, when the tool fails to find a model it may be because the specification is not satisfiable within the given scope, not that it is inconsistent in itself. However, the scope problem is orthogonal to the shortcoming addressed in this paper and it is assumed in this paper that AA never fails to find a model because of an insufficiently large scope. 
\vspace*{-5pt}
\section {Loy}\label{sec:loy}
\vspace*{-5pt}
Loy is a lightweight formal specification language for object-oriented programs. Its syntax is text-based and borrows keywords from common object-oriented program specification languages and its semantics is based on common notions of invariant and method specification \cite{leavens98}. Loy supports the specification of class interfaces within a single-inheritance hierarchy and includes the notions of subclass and subtype. Variables declared in a class specification are abstract specification variables \cite{cheon04} and methods are specified with preconditions, postconditions and frame conditions in a \emph{design-by-contract} manner \cite{meyer88}. Loy also includes \emph{depends clauses} to allow sound modular specification of the frame conditions \cite{muller03}. The syntax of a Loy specification is given in Figure~\ref{fig:abssyntax} ($c$, $v$, $m$ and $\phi$ respectively denote a class name, a variable name, a method name and set of first-order formulae; \emph{token}+ denotes one or more instances of \emph{token} and \emph{[ token ]} denotes zero or one instances of \emph{token}). The main constructs of Loy are introduced below through three example class specifications.
\vspace*{-5pt}
\begin{figure}[htp]
\framebox[\textwidth]{
\begin{tabular}{rcll}
\nterm{spec} & ::= & \nterm{class}+ & \hspace*{8pt}specification\\
\nterm{cspec} & ::= & \term{class} $c_1$ \term{ext} $c_2$ \nterm{body} \nterm{mspec}+ & \hspace*{8pt}class specification\\
\nterm{body} & ::= & \nterm{dec}+ \nterm{dep}+ \nterm{inv}+ & \hspace*{8pt}class body\\
\nterm{dec} & ::= & $v$ \textbf{:} $c$ & \hspace*{8pt}field declaration\\
\nterm{dep} & ::= &  \term{depends} $v_1$ \term{<-} $v_2 \ldots$ $v_n$ & \hspace*{8pt}depends clause\\
\nterm{inv} & ::= & \term{invariant} $\phi_I$ & \hspace*{8pt}invariant \\
\nterm{mspec} & ::= & \emph{[} $c$ \emph{]} $m$ \nterm{dec}+  \emph{[} \nterm{pre} \emph{]}\emph{[} \nterm{post} \emph{]}\emph{[} \nterm{mod} \emph{]}& \hspace*{8pt}method specification\\
\nterm{pre} & ::= & \term{requires} $\phi_P$ +& \hspace*{8pt}precondition \\
\nterm{post} & ::= & \term{ensures} $\phi_Q$ +& \hspace*{8pt}postcondition \\
\nterm{mod} & ::= & \term{modifies} $v_1 \ldots v_n$ & \hspace*{8pt}frame condition \\
\end{tabular}
}
\caption{Abstract syntax for a Loy specification}\label{fig:abssyntax}
\end{figure}
\begin{minipage}{6.05cm}
\begin{example}(Project.loy)\label{ex:project}\\
\term{class Project \{ }
\\
\hspace*{20pt}\term{manager : Manager }
\\	
\hspace*{20pt}\term{invariant some manager} 
\\
\term{\}}
\end{example}
\begin{example}(Employee.loy)\label{ex:employee}\\
\term{class Employee \{ }
\\
\hspace*{20pt}\term{project : Project}
\\	
\hspace*{20pt}\term{invariant no project.manager}
\\
\term{\\}
\hspace*{20pt}\term{assign (p : Project)}
\\
\hspace*{40pt}\term{requires no project}
\\
\hspace*{40pt}\term{ensures project' = p}
\\
\hspace*{40pt}\term{modifies project} 
\\
\term{\}}
\end{example}
\end{minipage}
\begin{minipage}{6.15cm}
\begin{example}(ManagedEmployee.loy)\label{ex:managed}\\
\term{class ManagedEmployee ext Employee \{ }
\\
\hspace*{20pt}\term{manager : Manager}
\\
\hspace*{20pt}\term{depends manager <- project}
\\
\term{\\}
\hspace*{20pt}\term{assign (p : Project)}
\\
\hspace*{40pt}\term{requires no project}
\\
\hspace*{40pt}\term{ensures project' = p}
\\
\hspace*{40pt}\term{ensures manager' = p.manager}
\\
\hspace*{40pt}\term{modifies project} 
\\
\term{\}}
\end{example}
\end{minipage}
 \vspace*{5pt}\\
\hspace*{\parindent} 
The above class specifications describe aspects of a relationship between employees, projects and managers (the specification for the class \emph{Manager} is not shown). A class \emph{Project} (Example~\ref{ex:project}) has a field \emph{manager} of type \emph{Manager} that represents the manager of the project and an invariant that specifies that \emph{manager} must not be null (\emph{some manager}). The declarations and invariants constitute the core constraints of a Loy specification. Formulae are written in a first-order logic that includes the logical connectives \emph{and}, \emph{or}, \emph{implies}, \emph{not}, the quantifiers \emph{all} and \emph{exists} and the Alloy keywords \emph{no} and \emph{some} to specify that a set is empty and nonempty, respectively. A class \emph{Employee} (Example~\ref{ex:employee}) has a field \emph{project} of type \emph{Project} and an invariant that states that the employee's project has no manager (\emph{no project.manager)}. A method \emph{assign}, which takes a parameter of type \emph{Project}, is specified: an employee can be assigned to a project only if no project has been assigned already. This constraint (given by the \emph{requires} clauses) is the precondition of the method. The postcondition of the method (given by the \emph{ensures} clause) simply states that the project \emph{p} is assigned to the field \emph{project} in the after state of the method. A field reference with a prime (e.g. \emph{project $'$}) denotes the value of the field in the after state of a method. Field references appearing in method specifications without a prime denote values in the before state. The frame condition of the method (given by the \emph{modifies} clause) states that only the value of the field \emph{project} may be affected by an invocation of \emph{assign}.
\vspace*{0pt}\\
\hspace*{\parindent} 
\emph{Employee} is extended by a class \emph{ManagedEmployee} (Example~\ref{ex:managed}), which adds a field \emph{manager} of type \emph{Manager} to the field \emph{project} inherited from \emph{Employee}. However, the specification for \emph{assign} is not inherited but overridden. The postcondition of the method is strengthened and the precondition and frame condition are unchanged. The postcondition now states that project \emph{p} is assigned to \emph{project} and the manager \emph{p.manager} is assigned to \emph{manager} in the after state of the method. The depends clause states that when the field \emph{project} changes value, the field \emph{manager} may also change value. Thus, the overriding version of \emph{assign} may change the value of \emph{manager} and yet have an equivalent frame condition to the overridden method, as the behavioural subtype relation requires \cite{liskov94}.
\vspace*{0pt}\\
\hspace*{\parindent} 
Loy borrows features from both Alloy and JML. From Alloy it takes its first-order relational logic and from JML it takes its modular object-oriented specification structure. But, unlike Alloy, the concept of state in a Loy specification is semantically implicit and the syntactic constructs of the language are tailored to the specification of object-oriented programs. For example, we can simply denote the before and after state value of the field \emph{manager} by writing \emph{manager} and \emph{manager~$'$}. Further, mathematical constructs implicit in the declarations of Alloy fields do not appear in Loy, which requires all constraints to be expressed explicitly as invariants. This allows the consistency of these constraints to be more easily checked against the trivially satisfiable field declarations and depends clauses. For example, a field \emph{f} in Alloy might be declared as \emph{f : F} or \emph{f : lone F}, indicating that \emph{f} is nonempty or possibly empty, respectively. We can express that a field \emph{f} is nonempty in Loy with the invariant \emph{some f}. Similarly, while Loy allows only declarations of the form \emph{f : F} and \emph{f : set F}, Alloy allows binary relation types (\emph{f : F -> G}) and set expression types (\emph{f : F + G}, for union; \emph{f : F - G}, for difference; and \emph{f : F \& G}, for intersection). In Loy, such constructs may be represented explicitly with user-defined data types. Finally, unlike JML, Loy specifications do not declare program variables. All variables of a Loy specification are abstract and no means of relating abstract variables to program variables is provided (c.f. \emph{represents} clauses in JML \cite{leavens98} and Spec$\sharp$ \cite{barnett04}). This avoids the sometimes confusing distinction between abstract and program variables in a specification and the need to declare program variables with one access modifier for the implementation and another for the specification (c.f. the \emph{spec\_public} access modifier in JML \cite{leavens98}).
\vspace*{0pt}\\
\hspace*{\parindent} 
Loy specifications can be analysed using AA through an encoding of Loy into Alloy,  presented in Section~\ref{sec:encoding}. For example, the satisfiability of the specification of the method \emph{assign} in \emph{Employee} can be checked using AA by checking whether the Alloy encoding of the conjunction of the precondition, postcondition and frame condition
is consistent with respect to the core constraints. The tool searches for a model of the specification that satisfies this conjunction and, failing to find one, suggests that the specification of the method may be inconsist\-ent. What is wrong with the specification of \emph{assign}? Of course, nothing is wrong with the specification of the method: the inconsistency lies in the specification of the invariants of the classes \emph{Project} and \emph{Employee}. The invariant for \emph{Project} states that there must be a manager while the invariant for \emph{Employee} states that its project has no manager.  No model that satisfies the above property is found because there is no model that satisfies the specification of \emph{Project} and \emph{Employee}. But AA does not point a user in this direction. 
\vspace*{-5pt}
\section {Patterns of Analysis}\label{sec:patterns}
\begin {figure}[htb]
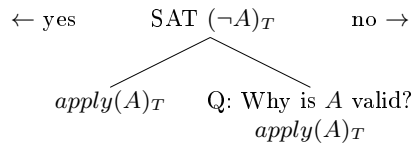

\hspace*{.8cm}
\small{
\Tree [.{$\leftarrow$ yes\hspace*{1cm}SAT $(\lnot A)_T$\hspace*{1cm}no $\rightarrow$} {$apply(A)_T$} {Q: Why is $A$ valid?\\$apply(A)_T$} !\faketreewidth{blahblahblahblahblahblah} ] 
}
\caption {Pattern 1: Negation}\label{fig:not}
\end {figure}

\begin {figure}[htb]
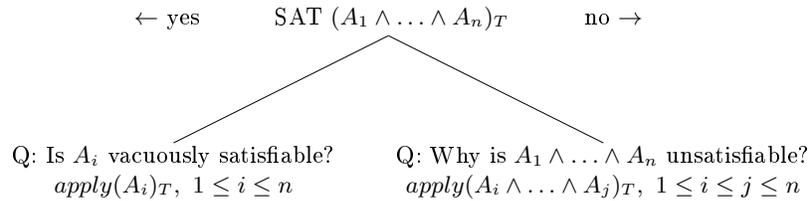

\hspace*{.55cm}
\small{
\Tree [.{$\leftarrow$ yes\hspace*{1cm}SAT $(A_1 \land \ldots \land A_n)_T$\hspace*{1cm}no $\rightarrow$} {Q: Is $A_i$ vacuously satisfiable?\\$apply (A_i)_T,~1 \leq i \leq n$} !\faketreewidth{blahblahblahblahblahblahblblahblahblah} {Q: Why is $A_1 \land \ldots \land A_n$ unsatisfiable?\\$apply (A_i \land \ldots \land A_j)_T,~1 \leq i \leq j \leq n$}  ] 
}
\caption {Pattern 2: Conjunction}\label{fig:con}
\end {figure}

\begin {figure}[htb]
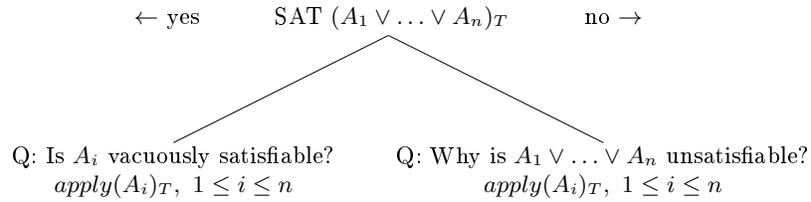

\hspace*{.55cm}
\small{
\Tree [.{$\leftarrow$ yes\hspace*{1cm}SAT $(A_1 \lor \ldots \lor A_n)_T$\hspace*{1cm}no $\rightarrow$} {Q: Is $A_i$ vacuously satisfiable?\\$apply (A_i)_T,~1 \leq i \leq n$} !\faketreewidth{blahblahblahblahblahblahblblahblahblah} {Q: Why is $A_1 \lor \ldots \lor A_n$ unsatisfiable?\\$apply (A_i)_T,~1 \leq i \leq n$}  ] 
}
\caption {Pattern 3: Disjunction}\label{fig:dis}
\end {figure}

\begin {figure}[ht]
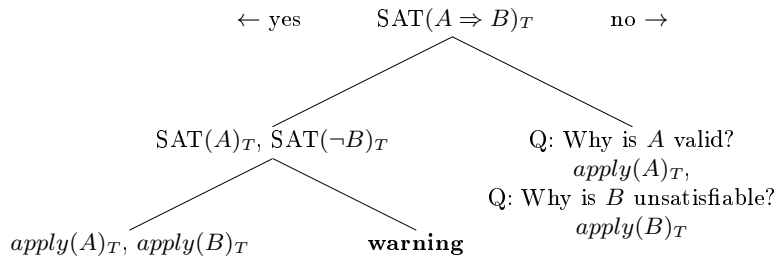

\hspace*{.7cm}
\small{
\Tree [.{$\leftarrow$ yes\hspace*{1cm}SAT$(A \Rightarrow B)_T$\hspace*{1cm}no $\rightarrow$} [.{SAT$(A)_T$, SAT$(\lnot B)_T$} {$apply (A)_T$, $apply(B)_T$} \olabel{warning} !\faketreewidth{blahblahblahblahblahblahblah} ] {Q: Why is $A$ valid?\\$apply(A)_T$,\\Q: Why is $B$ unsatisfiable?\\$apply(B)_T$} !\faketreewidth{blahblahblahblahblahblahblah} ]
}
\caption {Pattern 4: Implication}\label{fig:implication}
\end {figure}

\begin {figure}[ht]
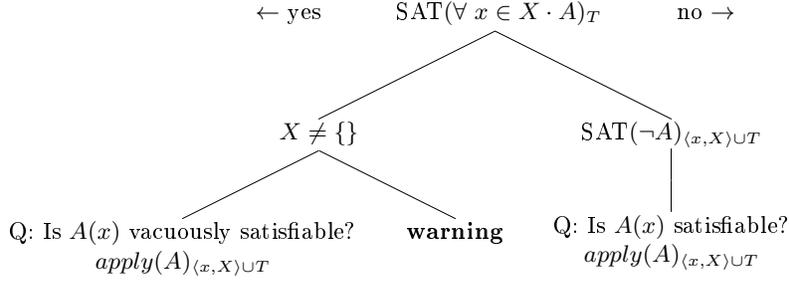

\hspace*{.5cm}
\small{
\Tree [.{$\leftarrow$ yes\hspace*{1cm}SAT$(\forall~x \in X \cdot A)_T$\hspace*{1cm}no $\rightarrow$} [.{$X \neq \{\}$} {Q: Is $A(x)$ vacuously satisfiable?\\$apply(A)_{\langle x, X \rangle \cup T}$} {\olabel{warning}} !\faketreewidth{blahblahblahblah} ]  {SAT$(\lnot A)_{\langle x, X \rangle \cup T}$\\\large{|}\\\large{|}\\Q: Is $A(x)$ satisfiable?\\$apply(A)_{\langle x, X \rangle \cup T}$} !\faketreewidth{blahblahblahblahblahblahblah} ]
}
\caption {Pattern 5: Universal Quantification}\label{fig:univ}
\end {figure}

\begin {figure}[ht]
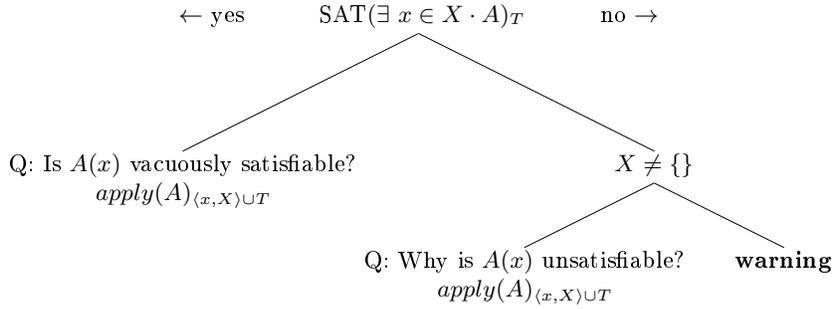

\hspace*{.2cm}
\small{
\Tree [.{$\leftarrow$ yes\hspace*{1cm}SAT$(\exists~x \in X \cdot A)_T$\hspace*{1cm}no $\rightarrow$} {Q: Is $A(x)$ vacuously satisfiable?\\$apply(A)_{\langle x, X \rangle \cup T}$}
[.{$X \neq \{\}$} {Q: Why is $A(x)$ unsatisfiable?\\$apply(A)_{\langle x, X \rangle \cup T}$} {\olabel{warning}} !\faketreewidth{blahblahblahblah} ] ]
}
\caption {Pattern 6: Existential Quantification}\label{fig:exist}
\end {figure}

\vspace*{-5pt}
This Section presents some \emph{patterns of analysis} --- one for each of the logical connectives and quantifiers of Loy --- that may be applied to Loy specifications and automated with AA to address the shortcoming illustrated above. The patterns are presented in Figures~\ref{fig:not}--\ref{fig:exist} as decision trees in which a non-terminal node represents a satisfiability query submitted to AA with respect to a given specification. Note that it is only necessary to consider satisfiability queries because validity of a formula $A$ is handled by checking the satisfiability of $\lnot A$. If there is no model of the specification that satisfies $\lnot A$ the tool informs the user that $A$ is possibly valid.
 \vspace*{0pt}\\
\hspace*{\parindent} 
It is assumed that the consistency of the core constraints of a specification is established before checking other properties, such as the correctness of method specifications. The consistency of the core constraints of a specification are therefore checked first by checking their consistency with $\top$. If the core constraints of a specification are inconsistent, further investigation is carried out by applying the patterns to each invariant in turn, checking its consistency with respect to the declarations of the specification and in conjunction with the other invariants. 
 \vspace*{-12pt}\\
\hspace*{\parindent} 
Given a specification $S$, SAT$(A)_T$ represents a query that asks whether there is a model of the core constraints of $S$ that satisfies $A$, where any free variables in $A$ are bound according to the type information held in the set $T$. At the start of an application of the patterns, $T$ is empty but the patterns for universal and existential quantification (Figures~\ref{fig:univ} and \ref{fig:exist}) strip off the outer quantifier leaving the bound variable free. For example, if the pattern for universal quantification is applied to the formula $\forall~x \in X \cdot A(x)$, the quantification $\forall~x \in X$ is stripped off before applying the pattern for $A(x)$ and the information that $x$ is of type $X$ is recorded in $T$. It is also assumed that the declarations of a specification provide contextual information for the specification variables in $A$. For ex\-ample, checking whether the precondition of \emph{assign} (i.e. \emph{no projects}) is satisfiable, involves checking whether there is an instance of \emph{Employee}, \emph{e} say, such that \emph{no e.projects} is satisfied. The information that \emph{projects} is a field of \emph{Employee} comes from the declarations of the specification under analysis. 
 \vspace*{0pt}\\
\hspace*{\parindent} 
From a non-terminal node, the left branch indicates that the formula is satisfiable and the right branch indicates that the formula is unsatisfiable. Terminal nodes represent one of two possible outcomes: 
\begin{itemize}
\item \olabel{warning} signals that the formula of the root node may be vacuously satisfiable or vacuously unsatisfiable and the feedback to the original query is possibly misleading; 
\item $apply (A)_T$ indicates that the pattern for formula $A$ should be applied where further diagnosis is required ($T$ carries type details for free variables in $A$).
\end{itemize}
A question annotating a terminal node (e.g. \emph{Q: Why is A valid?}) gives the user a context for the further application of patterns. 
Feedback from each query along a path through the patterns is noted, so that as much diagnostic information as possible is gathered for the original query. AA provides instantiations of a specification each time a query is satisfiable, i.e. each left branch of a tree produces an assignment for the variables of the specification and the queried formula. 
 \vspace*{0pt}\\
\hspace*{\parindent} 
\textbf{Atomic expressions.} An atomic expression is one of the two base cases (the other being the issue of a warning) in the application of a pattern. An atomic expression in Loy is a well-formed boolean expression that contains no logical connective. If we check the satisfiability of an atomic expression with AA we will be provided with an instantiation of the specification that satisfies the atomic expression, if the expression is satisfiable, and nothing if the expression is unsatisfiable. As long as the core constraints of the specification are consistent there is little scope for vacuity in the feedback to such a query. But whether an atomic expression is unsatisfiable, valid or neither can be discovered by testing the negation of the expression. For example, if $p$ is satisfiable and $\lnot p$ is unsatisfiable, then $p$ is valid. 
 \vspace*{0pt}\\
\hspace*{\parindent} 
\textbf{Pattern 1: Negation.} 
(Figure~\ref{fig:not}.) The pattern for negation allows the val\-idity of a formula to be queried by checking whether the  negation of the formula is satisfiable. For example, to check whether $A$ is valid, the satisfiability of $\lnot A$ is checked. If $\lnot A$ is satisfiable, $A$ is not valid. But $\lnot A$ may be vacuously satisfiable, so the satisfiability of $A$ should be checked by applying the pattern for $A$. On the other hand, if $\lnot A$ is unsatisfiable, $A$ is valid. But the validity of $A$ can be further investigated by applying its pattern. 
 
\textbf{Pattern 2: Conjunction.} (Figure~\ref{fig:con}.) The pattern for conjunction decomposes a formula into its conjuncts to identify either whether the formula is vacuously satisfiable or, if it is unsatisfiable, which conjuncts contain inconsistencies. For example, when checking the consistency of a method specification, the conjunction of the precondition, postcondition and frame condition is checked against the core constraints. Here, applying this pattern could uncover a vacuously satisfiable (and hence useless) postcondition or identify which part of the method specification contains an inconsistency. The pattern is applied by checking the satisfiability of a conjunction $A_1 \land \ldots \land A_n$. If $A_1 \land \ldots \land A_n$ is satisfiable, each conjunct is satisfiable. But $A_1 \land \ldots \land A_n$ may be vacuously satisfiable, so the pattern for each conjunct $A_i$ is applied since, if $A_i$ is vacuously satisfiable, so is $A_1 \land \ldots \land A_n$. On the other hand, if $A_1 \land \ldots \land A_n$ is unsatisfiable, at least one conjunct is unsatisfiable. Which conjunct or conjuncts are unsatisfiable could be established by applying the patterns for all combinations of conjunct $A_i \land \ldots \land A_j,~1 \leq i \leq j \leq n$. 
 
\textbf{Pattern 3: Disjunction.} (Figure~\ref{fig:dis}). The pattern for disjunction decomposes a formula into its disjuncts to identify either whether the formula is vacuously satisfiable or, if it is unsatisfiable, why all disjuncts are unsatisfiable. The pattern is applied by checking the satisfiability of a disjunction $A_1 \lor \ldots \lor A_n$. If $A_1 \lor \ldots \lor A_n$ is satisfiable, at least one disjunct is satisfiable. But, as for Pattern~2, it is established whether $A_1 \lor \ldots \lor A_n$ is vacuously satisfiable by app\-lying the pattern for each disjunct $A_i$ since, if $A_i$ is vacuously satisfiable, so is $A_1 \lor \ldots \lor A_n$. On the other hand, if $A_1 \lor \ldots \lor A_n$ is unsatisfiable, no disjunct is satisfiable. The pattern for each $A_i$ is applied to investigate its unsatisfiability. 
 
\textbf{Pattern 4: Implication.} (Figure~\ref{fig:implication}.) The pattern for implication exposes whether or not an implication is vacuously satisfiable because of an unsatisfiable antecedent or valid consequent. For example, if $A$ is the precondition for a method and $B$ the postcondition, we might ask whether $A \Rightarrow B$ for all models of the specification. If the precondition is unsatisfiable the tool will suggest that $A \Rightarrow B$ may be valid because it cannot satisfy $\lnot (A \Rightarrow B)$. The pattern is applied by checking the satisfiability of an implication $A \Rightarrow B$. If $A \Rightarrow B$ is satisfiable, the satisfiability of both $A$ and $\lnot B$ is checked. If either $A$ or $\lnot B$ is unsatisfiable, $A \Rightarrow B$ is vacuously satisfied and a warning is issued. Otherwise, the subformulae $A$ and $B$ may be further investigated by applying their patterns. On the other hand, if $A \Rightarrow B$ is unsatisfiable, $A$ must be valid and $B$ must be unsatisfiable. Again, the subformulae may be investigated further by applying their patterns. 
 
\textbf{Pattern 5: Universal Quantification.} (Figure~\ref{fig:univ}.) 
Since the logic of Loy (and Alloy) is sorted, a universally quantified formula may be vacuously satisfiable because the domain ranged over by the quantifier is empty. 
For example, if a class specification $C$ is inconsistent, then the formula $\forall~c \in C \cdot A$ will be vacuously satisfied because $C$ is unsatisfiable and hence there are no instances $c$. The pattern is applied by checking the satisfiability of a universally quantified formula $\forall~x \in X \cdot A$. If $\forall~x \in X \cdot A$ is satisfiable, the domain $X$ is checked to see if it is empty. If it is, a warning is issued because $\forall~x \in X \cdot A$ is vacuously satisfiable. If the domain is not empty, the satisfiability of $A$ for all $x \in X$ is checked by applying the pattern for $A (x)$ for an arbitrary $x$ (adding the binding $\langle x, X \rangle$ to $T$). On the other hand, if $\forall~x \in X \cdot A$ is unsatisfiable, it should be established for what assignments to $x$ $A$ is unsatisfied. $\lnot A$ is satisfied for at least one value of $x$ but its satisfiability is checked (adding the binding $\langle x, X \rangle$ to $T$) simply to acquire a model of the specification that contains a counterexample to $\forall~x \in X \cdot A$. The satisfiability of $A$ for any assignment to $x$ can be investigated further by applying the pattern for $A$ (adding the binding $\langle x, X \rangle$ to $T$).  

\textbf{Pattern 6: Existential Quantification.} (See Figure~\ref{fig:exist}.) An existentially quantified formula may be vacuously unsatisfiable because the domain ranged over by the quantifier is empty. 
For example, similarly to the pattern for universal quantification, if a class specification $C$ is inconsistent, then the formula $\exists~c \in C \cdot A$ will be vacuously unsatisfied because $C$ is unsatisfiable and hence there are no instances $c$. The pattern is applied by checking the satisfiability of an existentially quantified formula $\exists~x \in X \cdot A$. If $\exists~x \in X \cdot A$ is satisfiable, it is investigated whether $A$ is vacuously satisfiable for an assignment to $x$ by applying the pattern for $A (x)$ for an arbitrary $x$ (adding the binding $\langle x, X \rangle$ to $T$). On the other hand, if $\exists~x \in X \cdot A$ is unsatisfiable, the domain $X$ is checked to see if it is empty. If it is, a warning is issued because $\exists~x \in X \cdot A$ is vacuously unsatisfiable. If the domain is not empty, the satisfiability of $A$ for all assignments to $x$ is checked by applying the pattern for $A$  (adding the binding $\langle x, X \rangle$ to $T$). 
 
\textbf{Examples.} This Section ends with two examples. Let $S$ be a specification, $p$, $q$ and $r$ be atomic expressions and $C$ be a consistent class specification in $S$ (i.e. $C$ is a nonempty domain when used as a logical sort). Let the formula
$\forall~c : C \cdot P(c)~\Longrightarrow~(Q(c)~\lor~R(c))$
be an invariant of $C$. With respect to a specification $S$, $P(c)$ is unsatisfiable, $Q(c)$ is valid and $R(c)$ is satisfied in some models of $S$ and unsatisfied in others. The formula is satisfiable because $C$ is consistent but it should be established whether or not it is vacuously satisfiable. Application of the pattern for this formula follows the path shown below.
\vspace*{8pt}\\
\small
\hspace*{20pt}Pattern 5 -- Universal quantification.\\
\hspace*{20pt}$\forall~c : C \cdot P(c)~\Rightarrow~(Q(c)~\lor~R(c))$ .. satisfiable \\
\hspace*{20pt}| \\
\hspace*{20pt}$C \neq \{\}$  .. satisfiable \\
\hspace*{20pt}| \\
\hspace*{20pt}Pattern 4 -- Implication.  \\
\hspace*{20pt}$P(c)~\Rightarrow~(Q(c)~\lor~R(c))$  .. satisfiable, $\langle c, C \rangle$\\
\hspace*{20pt}| \\
\hspace*{20pt}$P(c)$  .. unsatisfiable, $\langle c, C \rangle$ \\
\hspace*{20pt}| \\
\hspace*{20pt}\olabel{warning} (unsatisfiable antecedent)  \\
\vspace*{-4pt}\\
\normalsize
The invariant of $C$ is entailed by the specification simply because $P(c)$ is unsatisfiable. But if, in fact,  $P(c)$ was wrongly assumed to be satisfied in some models of $S$, the original feedback from AA, which simply suggested the property may be valid, would be inadequate. 
 
To give an applied example, the patterns can be used to question the feedback given when checking the satisfiability of the specification of the method \emph{assign} in the class specification for \emph{Employee} (see Example~\ref{ex:employee}, Section~\ref{sec:loy}). The satisfiability of the method specification is checked by checking the satisfiability of the precondition, postcondition and frame condition with respect to the class specifications for \emph{Employee} and \emph{Project}, which is the type of the parameter \emph{p}. For brevity, let $P \land Q \land R$ denote the conjunction of the precondition, postcondition and frame condition. The application of the pattern for this formula (the pattern for existential quantification) follows the short path shown below, which terminates with a warning that the formula is vacuously unsatisfiable. 
\vspace*{8pt}\\
\small
\hspace*{20pt}Pattern 6 -- Existential quantification. \\
\hspace*{20pt}$\exists~e : Employee~\exists~p : Project \cdot P \land Q \land R$ .. unsatisfiable \\
\hspace*{20pt}| \\
\hspace*{20pt}$Employee \neq \{\}$ .. unsatisfiable \\
\hspace*{20pt}| \\
\hspace*{20pt}\olabel{warning} (specification for $Employee$ is unsatisfiable)  
\vspace*{8pt}\\
\normalsize
The domain \emph{Employee} is empty because the specification of \emph{Employee} contains an inconsistency and cannot be instantiated. However, the clash between the two invariants is noted and the invariant for \emph{Project} is removed. The query is run a second time and AA suggests that the property is consistent and provides an instantiation of the specification and an assignment to the variables $e$ and $p$.
\vspace*{-5pt}
\section {Encoding Loy in Alloy}\label{sec:encoding}
\vspace*{-5pt}
This Section illustrates the encoding of Loy in Alloy. The encoding both gives Loy a semantics and allows automation of the patterns to be implemented on top of AA. We define the encoding of Loy in Alloy through the \emph{encoding function} 
\[
\textrm{\tv} :~\mathcal{L}{_\textrm{\small Loy}} \rightarrow \mathcal{L}{_\textrm{\small Alloy}}
\]
that maps a Loy specification to an Alloy specification such that the models of a Loy specification are the models of that specification encoded in Alloy. That is, for a Loy specification $s$, \textsf{mod} ($s$) $=$ \textsf{mod} (\tv($s$)). The meaning of a Loy specification is thus the meaning of the Alloy specification that encodes it. The formal definition of the encoding function is omitted from this paper but the main features of the encoding are introduced below using the encodings of \emph{Emp\-loyee} and \emph{ManagedEmployee}. All encodings are generated automatically.
\vspace*{0pt}\\
\hspace*{\parindent}A Loy class specification is encoded as an Alloy signature paragraph and set of predicate paragraphs, one for each invariant, precondition, postcondition and frame condition of the class specification. The encoding of the field declaration and depends clause of \emph{ManagedEmployee} (see Example~\ref{ex:managed}) is shown below. 
\vspace{8pt}
\\
\hspace*{20pt}\term{sig ManagedEmployee extends \{ manager : lone Manager \} \{}\\
\hspace*{40pt}\term{// depends manager <- project}\\
\hspace*{40pt}\term{idxOf (fields, manager) -> idxOf (fields, project) in depends}\\
\hspace*{40pt}\term{\# depends = 1}\\
\hspace*{20pt}\term{\}}\\
\hspace*{20pt}\term{fact ManagedEmployee\_fieldtable \{}\\
\hspace*{40pt}\term{let idx0 = Ord\_/Ord.first, idx1 = Ord\_/next (idx0) \{}\\
\hspace*{60pt}\term{all o : ManagedEmployee \{}\\
\hspace*{80pt}\term{at (o.fields, idx0) = o.manager}\\
\hspace*{80pt}\term{at (o.fields, idx1) = o.project} \\ 
\hspace*{20pt}\term{\}\}\}}
\vspace{8pt}
\\
The field declaration of the class specification is encoded as a field declaration of the signature paragraph. For example, the declaration  \emph{manager : Manager} in the Loy specification becomes the declaration \emph{manager : lone Manager} in a signature paragraph \emph{ManagedEmployee}. The modifier \emph{lone} indicates that the field \emph{projects} may be empty, capturing the default semantics of the Loy declaration. Every signature that encodes a class specification inherits the fields \emph{fields} and \emph{depends} from a root signature \emph{Obj}, shown below. 
\vspace{8pt}
\\
\hspace*{20pt}\term{sig Obj \{ fields : Seq [Obj], depends : SeqIdx -> SeqIdx \}}
\vspace{8pt}
\\
The field \emph{fields} is a sequence representing the fields of the class specification encoded by a signature that allows us to refer to the index of a field, i.e. the \emph{location} of a field in a class rather than its value for a particular instance. The fact paragraph \emph{ManagedEmployee\_fieldtable} represents the field table (an Alloy sequence) for \emph{ManagedEmployee}. The field \emph{depends} is a binary relation that represents the depends relation of the class specification encoded by a signature.  A depends clause is encoded as a constraint that specifies that \emph{depends} contains pairings of the field on the left of the arrow with each of the fields on the right. For example, \emph{depends manager <- project} is encoded as the constraint that the index of \emph{manager} and the index of \emph{project} are a pair in the \emph{depends} field of \emph{ManagedEmployee}. Further, since it is known that no class specification extends \emph{ManagedEmployee} for this encoding and, therefore, that the depends relation will not be added to, its cardinality is declared to be 1 (\emph{\# depends = 1}) to prevent erroneous extra pairs being inserted by AA during analysis.
\vspace*{0pt}\\
\hspace*{\parindent}
An invariant of a class specification is encoded as a predicate paragraph containing a formula that applies the invariant to all instances of the signature that encodes the class specification. For example, \emph{ManagedEmployee} inherits from \emph{Employee} the invariant \emph{no project.manager} (see Example~\ref{ex:employee}), which is encoded as the predicate paragraph shown below.
\vspace{8pt}
\\
\hspace*{20pt}\term{pred Employee\_I () \{ all x : Employee | no x.project.manager \}}
\vspace{8pt}
\\
Encoding an invariant as a predicate paragraph allows analyses of a specification to be carried out with and without the invariant present. Specifically, this allows us to check the consistency of the invariant itself with respect to the rest of a specification.
\vspace*{0pt}\\
\hspace*{\parindent}
To represent the concept of state implicit in a Loy specification the signature paragraphs \emph{Id} and \emph{State} are created. The \emph{Id} and \emph{State} signatures for the encoding of our example specification are shown below. 
\vspace{8pt}
\\
\hspace*{20pt}\term{sig Id \{ \}}\\
\hspace*{20pt}\term{sig State \{}\\
\hspace*{40pt}\term{manager : Id lone -> lone Manager,}\\
\hspace*{40pt}\term{project : Id lone -> lone Project,}\\
\hspace*{40pt}\term{employee : Id lone -> lone Employee,}\\
\hspace*{40pt}\term{managedEmployee : Id lone -> lone ManagedEmployee}\\ 
\hspace*{20pt}\term{\}}
\vspace{8pt}
\\
The fields of the signature \emph{State} are partial mappings from a set of references (\emph{Id}) to sets of instances of the signatures that encode the class specifications. These mappings represent persistant references to instances of a class specification across states. For example, given two states, \emph{s} and \emph{s~$'$}, and an \emph{Id}, \emph{i}, we can refer to an instance of \emph{ManagedEmployee} in \emph{s} and \emph{s~$'$}, respectively, with the expressions \emph{s.managedEmployee[i]} and \emph{s~$'$.managedEmployee[i]} where \emph{i} provides a reference to an instance across states. Strictly, \emph{s.managedEmployee[i]} and \emph{s~$'$.managedEmployee[i]} denote different instances of the Alloy signature \emph{ManagedEmployee} that represent (possibly) different values of a single instance of the Loy class specification \emph{ManagedEmployee}.
\vspace*{0pt}\\
\hspace*{\parindent}
A method specification is encoded as three predicate paragraphs, one each for the precondition (given a $P$ suffix), postcondition (given a $Q$ suffix) and frame condition (given an $F$ suffix). The encoding of the method specification for \emph{assign} in \emph{ManagedEmployee} is shown below.
\vspace{8pt}
\\
\hspace*{20pt}\term{pred ManagedEmployee\_assign\_P (i : Id, s0 : State, p : Project) \{}\\
\hspace*{40pt}\term{no s0.managedEmployee[i].project}\\
\hspace*{20pt}\term{\}}\\
\hspace*{20pt}\term{pred ManagedEmployee\_assign\_Q (i : Id, s0, s1 : State, p : Project) \{}\\
\hspace*{40pt}\term{s1.managedEmployee[i].project = p}\\
\hspace*{40pt}\term{s1.managedEmployee[i].manager = p.manager }\\
\hspace*{20pt}\term{\}}\\
\hspace*{20pt}\term{pred ManagedEmployee\_assign\_F (i : Id, s0, s1 : State, p : Project) \{}\\
\hspace*{40pt}\term{let o = s0.managedEmployee[i], o' = s1.managedEmployee[i] \{}\\
\hspace*{60pt}\term{all k : Seq/SeqIdx \{}\\
\hspace*{80pt}\term{at (o.fields, k) = at (o'.fields, k) ||}\\
\hspace*{80pt}\term{k = idxOf (o.fields, o.project) }\\
\hspace*{60pt}\term{\}}\\
\hspace*{60pt}\term{all x : ID \{}\\
\hspace*{80pt}\term{s1.managedEmployee[x] = s0.managedEmployee[x] || x = i}\\
\hspace*{80pt}\term{s1.manager[x] = s0.manager[x]}\\
\hspace*{80pt}\term{s1.project[x] = s0.project[x]}\\
\hspace*{80pt}\term{s1.employee[x] = s0.employee[x]} \\ 
\hspace*{20pt}\term{\}\}\}}
\vspace{8pt}
\\
Special parameters representing a reference (\emph{i}) and before and after states (\emph{s0} and \emph{s1}) are added to the parameter \emph{p} of \emph{assign}. Unprimed and primed variables of the Loy specification, respectively denoting before and after state values, are then encoded as expressions that refer to instances of a signature relative to \emph{s0} or \emph{s1} through the persistent reference \emph{i}. For example, \emph{project} and \emph{project~$'$} are encoded as \emph{s0.managedEmployee[i].project} and \emph{s1.managedEmployee[i].project}.
\vspace*{0pt}\\
\hspace*{\parindent}
A frame condition is constructed from a modifies clause. The mappings represented by the fields of \emph{s0} and \emph{s1} are specified to be the same except possibly where a field appears in the modifies clause, permitting the method to alter its value. The after state has to be constructed from the before state explicitly. Consider the variable expression $v_1 \cdot \ldots \cdot v_n$. If this expression appears in the modifies clause of a method specification, an invocation of the method is permitted to change the value of $v_n$. But, because of the value semantics of Alloy, if the value of $v_i$ changes (for $1 < i \leq n$), then the value of $v_{i-1}$ must also be updated. Further, for each $v_i$, the fields that are not permitted to change must be constrained to have the same value in \emph{s0} and \emph{s1}. Finally, the values of the fields of \emph{s1} are constrained: only the mappings for types of fields that are permitted to change may differ from those of \emph{s0}. For example, since \emph{assign} only changes the value of its receiver, which is an instance of \emph{ManagedEmployee}, only the mapping from \emph{Id} to \emph{ManagedEmployee} may differ between \emph{s0} and \emph{s1}.
 \vspace*{0pt}\\
\hspace*{\parindent}
The queries represented by the tree nodes in the patterns of analysis are implemented as Alloy queries to AA. The formulae are encoded in Alloy and wrapped in predicate paragraphs. The type information carried around in the set $T$ becomes the parameter declarations for this predicate paragraph. For example, the query SAT$(A \Rightarrow B)_{ \langle x, X \rangle, \langle y, Y \rangle }$ is encoded as the Alloy predicate paragraph
\vspace{8pt}
\\
\hspace*{20pt}\term{pred (x : X, y : Y) \{ A implies B \}}
\vspace{8pt}
\\
and submitted to the tool as a consistency query. The feedback from AA for each query is then interpreted by the user in terms of the Loy specification under analysis.
\vspace*{-5pt}
\section {Related Work}\label{sec:related}
\vspace*{-5pt}
Specification animation such as that supported by the ProB \cite{leuschel03}, Bogor \cite{robby03} and JML-TT-Animator \cite{jtt} tools is a simple way to check basic properties of specifications, as inconsistency may be caught quickly in the exploratory environment provided by an animator. However, as with test-driven development the onus is on the practitioner to direct the testing in directions that catch the errors. AA enhanced by patterns of analysis removes much of this onus and guides the practitioner through an exhaustive automated analysis of a specification. 
 \vspace*{0pt}\\
\hspace*{\parindent}
The Analysable Annotation Language (AAL) \cite{khurshid02} is a notation that borrows from both Alloy and JML and is designed to annotate Java programs with JML-like specifications. An encoding of AAL in Alloy allows these specifications to be checked by AA. However, AAL focuses more on the checking of an implementation against a specification and not on the preliminary analysis of the specification itself. Alternate encodings of state in Alloy are presented in \cite{khurshid04}, which encodes state in order to use Alloy to support software testing, and \cite{marinov02}, which introduces VAlloy, an extension of Alloy that adds virtual functions and a simple state model to the language. However, neither encoding gives a full treatment of specifications that handle frame conditions or depends clauses. 
 \vspace*{0pt}\\
\hspace*{\parindent}
Finally, a translation from B AMN to Alloy is presented in \cite{mikhailov01}. The B toolset provides limited support for automatically discharging proof obligations and many proofs have to be constructed interactively. By complementing a theorem prover with AA, using the latter to check the consistency or validity of a property before attempting a proof, valuable confidence that the property is in fact provable is gained. Though this is another example of AA being invoked to support specification analysis, our approach addresses the need for a lightweight technique that provides immediate returns to the software developer.
\vspace*{-5pt}
\section {Conclusion}\label{sec:conclusion}
\vspace*{-5pt}
In order to make a lightweight formal technique even more attractive to commercial software developers, we have defined a lightweight formal object-oriented specification language that is supported, through an encoding into Alloy, by AA. We have also identified patterns of analysis that guide a developer through the analysis of a specification to establish its satisfiability before implementation. 
\vspace*{0pt}\\
\hspace*{\parindent}
We have presented the patterns in the context of AA but the patterns are intended to be generally applicable in any analysis environment that depends on querying properties for satisfiability with respect to a model. 
Automatic generation of an Alloy encoding from a Loy specification has been implemented but a full implementation of the approach is still to be completed. The patterns will be implemented on top of AA such that a satisfiability query submitted to the tool for a given formula will immediately invoke the pattern for that formula so that misleading feedback is never given. 
\vspace*{0pt}\\
\hspace*{\parindent}
A current limitation of this approach is the time it takes to run several satisfiability queries in a row. 
Most desktop machines take a couple of minutes to compute a single pattern before feedback is given, even for an Alloy encoding of a couple of hundred lines. One way to shorten computation time would be to couple AA with a version control system in order to keep track of which parts of a specification were known to be consistent with each other and then only recheck the satisfiability of properties that have been affected by editing  since the last analysis. Once an initial check had been performed, this would greatly reduce the computation that needed to be done in the application of each pattern.
Further research will also include a detailed study of the complexity and completeness of the analysis patterns proposed.

\tiny
\bibliographystyle{plain}
\bibliography{phd}

\end {document}